\documentclass[journal]{IEEEtran}

\usepackage{graphicx,cite,epsfig}
\usepackage{amsmath} 
\usepackage{amssymb}  
\DeclareMathOperator *{\argmin}{argmin}
\usepackage{pifont}
\usepackage{stmaryrd}
\usepackage{bbding}
\usepackage{subfigure}
\usepackage{algorithm}
\usepackage{algorithmic}
\usepackage{array}
\usepackage{amsthm}
\usepackage{multirow}
\usepackage{url}
\usepackage{mathrsfs}
\usepackage{color, soul} 
\usepackage{mathrsfs}
\usepackage{epstopdf}

\newcommand{\tabincell}[2]{\begin{tabular}{@{}#1@{}}#2\end{tabular}}
\usepackage{booktabs}
\usepackage{threeparttable}
\usepackage{amssymb}

\begin{document}
\title{On the Construction of $G_N$-coset Codes for Parallel Decoding}
\author{\IEEEauthorblockN{Xianbin~Wang\IEEEauthorrefmark{1}, Huazi~Zhang\IEEEauthorrefmark{1}, Rong~Li\IEEEauthorrefmark{1}, Jiajie~Tong\IEEEauthorrefmark{1}, Yiqun~Ge\IEEEauthorrefmark{2}, Jun~Wang\IEEEauthorrefmark{1}} \\
\IEEEauthorblockA{\IEEEauthorblockA{\IEEEauthorrefmark{1}Hangzhou Research Center, Huawei Technologies, Hangzhou, China}\\
   \IEEEauthorblockA{\IEEEauthorrefmark{2}Ottawa Research Center, Huawei Technologies, Ottawa, Canada}\\
Emails: \{wangxianbin1,zhanghuazi,lirongone.li,tongjiajie,yiqun.ge,justin.wangjun\}@huawei.com}
}
\maketitle
\thispagestyle{empty}
\begin{abstract}
In this work, we propose a type of $G_N$-coset codes for parallel decoding.
The parallel decoder exploits two equivalent decoding graphs of $G_N$-coset codes.
For each decoding graph, the inner code part is composed of independent component codes to be decoded in parallel.
The extrinsic information of the code bits is obtained and iteratively exchanged between the two graphs until convergence.
Accordingly, we explore a heuristic and flexible code construction method (information set selection) for various information lengths and coding rates.
Compared to the previous successive cancellation algorithm, the parallel decoder avoids the \emph{serial} outer code processing and enjoys a higher degree of parallelism.
Furthermore, a flexible trade-off between performance and decoding latency can be achieved with three types of component decoders.
Simulation results demonstrate that the proposed encoder-decoder framework achieves comparable error correction performance to polar codes with a much lower decoding latency.
\end{abstract}

\section{Introduction}\label{section_introduction}
\subsection{Preliminary}
$G_N$-coset codes, as defined by Ar{\i}kan in \cite{ArikanPolar}, are a class of linear block codes with the generator matrix $G_N$.

$G_N$ is an $N \times N$ binary matrix defined as
\begin{equation}\small\label{me_all}
\begin{aligned}
G_N \triangleq F^{\otimes n}
  \end{aligned},
\end{equation}
in which $N=2^n$ and $F^{\otimes n}$ denotes the $n$-th Kronecker power of $F=[\begin{smallmatrix}
   1 & 0  \\
   1 & 1
  \end{smallmatrix} ]$.

The encoding process is
\begin{equation}\small\label{me_all1}
\begin{aligned}
x_1^N = u_1^NG_N,
\end{aligned}
\end{equation}
where $x_1^N \triangleq \{x_1,x_2,\cdots,x_N\}$ and $u_1^N \triangleq \{u_1,u_2,\cdots,u_N \}$ denote the code bit sequence and the information bit sequence respectively.

An $(N,K)$ $G_N$-coset code \cite{ArikanPolar} is defined by an information set $\mathcal{A}\subset\{1,2,...,N\}$, $|{\cal A}| = K$.
Its generator matrix $G_N(\mathcal{A})$ is composed of the rows indexed by $\mathcal{A}$ in $G_N$.
Thus (\ref{me_all1}) is rewritten as
\begin{equation}\small\label{me_all}
\begin{aligned}
x_1^N = u(\mathcal{A})G_N(\mathcal{A}),
\end{aligned}
\end{equation}
where $u(\mathcal{A})\triangleq\{u_i | i\in \mathcal{A}\}$.

The key to constructing $G_N$-coset codes is to properly determine an information set $\mathcal{A}$.
RM codes\cite{RM} and polar codes\cite{ArikanPolar}, two well-known examples of $G_N$-coset codes, determine $\mathcal{A}$ in terms of Hamming weight and sub-channel reliability, respectively, which are referred to as RM principle and polar principle.

Polar codes are the first capacity-achieving channel codes \cite{ArikanPolar}. RM codes are proved to achieve the binary erasure channel capacity under the maximum-a-posteriori (MAP) decoding algorithm\cite{RM}.
Both codes have been adopted for 5G control channel.
\label{fig_decoding}
\begin{figure}[]
	\centering
	\includegraphics[width=3.5in]{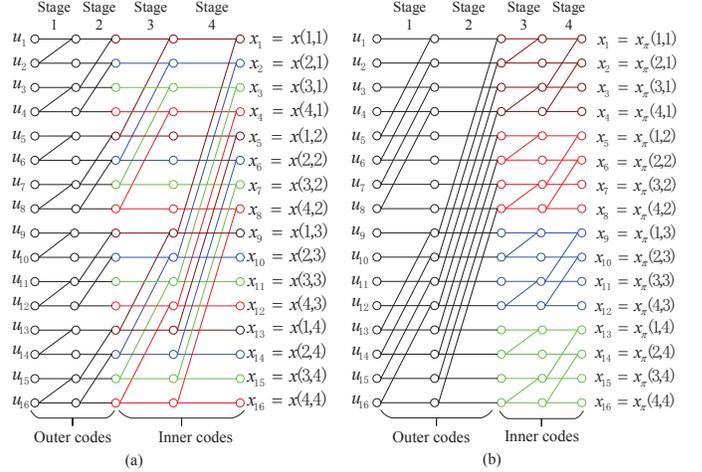}
	\caption{For $G_N$-coset codes, equivalent encoding graphs may be obtained based on stage permutations: (a) Ar{\i}kan's original encoding graph \cite{ArikanPolar} and (b) stage-permuted encoding graph.  Each node adds (mod-2) the signals on all incoming edges from the left and sends the result out on all edges to the right.  }
	\label{permutation}
\end{figure}

\subsection{Motivations and Contributions}
Both RM codes and polar codes are not designed for parallel decoding.
RM codes are only adopted for very short code lengths due to the lack of linear-complexity decoding algorithms.
Polar codes exhibit superior performance with successive cancellation (SC) based decoders.
But an SC decoder is \emph{serial} in nature\cite{ArikanPolar} as it requires $2N-2$ time steps for a length-$N$ code.

To seek parallelism on the decoding side, we propose a novel stage-permuted turbo-like decoding framework.
As shown in Fig.~\ref{permutation}(a), the encoding process of $G_N$-coset codes can be described by an $n$-stage encoding graph.
Therefore, $G_N$-coset codes can be considered as concatenation codes. The former and latter stages respectively correspond to outer and inner codes.
The inner code parts consist of \emph{independent} component codes that can be decoded in parallel (the $j$-th code bit of the $i$-th inner component code is denoted by $x(i,j)$ in Fig.~\ref{permutation}(a))~\cite{Syndrome}. In contrast, the outer code parts must be decoded successively, which is the major source of latency of all SC-based decoding algorithms.

Based on the above observation, the proposed algorithm improves decoding parallelism as follows.
First, equivalent encoding/decoding graphs (see Fig.~\ref{permutation}) of the same $G_N$-coset code can be obtained by permuting the encoding stages \cite{scpermutation1}.
Second, decoding is performed on each of these \emph{equivalent graphs}.
Within each graph, $\sqrt{N}$ inner component codes of length $\sqrt{N}$ are decoded in parallel, but the outer component codes are not processed.
Finally, decoding results from different graphs about the same code bit are exchanged to reach a consensus.
Fig.~\ref{permutation}(a) and Fig.~\ref{permutation}(b) show two \emph{equivalent graphs} for $N=16$. In each decoding graph, the inner code parts consist of $4$ component codes of length $4$.
Since only the inner code parts are decoded in parallel while the outer code processing is avoided, the proposed decoding algorithm exhibits a higher degree of parallelism.

Furthermore, we propose a new code construction principle (selection of $\cal A$) for the stage-permuted turbo-like decoding algorithm. In particular, we show that the principle to select $\cal A$ is to reduce the code rate of the inner codes. Accordingly, we explore a heuristic code construction that outperforms the RM and polar codes under the stage-permuted decoder.

\subsection{Related works}
In~\cite{STPC,NSTPC}, product codes with polar codes as component codes are studied, with the same aim of improving decoding parallelism.
As product codes, the codes are constructed from the component codes, which lead to a square number ($k^2$) of information bits.
In contrast, we follow Ar{\i}kan's $G_N$-coset code framework~\cite{ArikanPolar}, which is more general and flexible in two folds.
First, the code construction is defined directly by $\cal A$.
It naturally supports ``1-bit'' fine-granularity of information length.
Second, stage permutation potentially supports a more flexible framework with richer ($n!$ instead of two) combinations of outer-inner code concatenations.
Accordingly, iterative decoding can be performed on at most $n!$ stage-permuted graphs.

\section{Stage-permuted turbo-like decoding algorithm}\label{section_decoding}
The aforementioned stage-permuted turbo-like decoder is formally described in Algorithm~\ref{alg:stage-permute-decoder}.
\begin{algorithm}[htb]
\caption{A stage-permuted turbo-like decoder.}
\label{alg:stage-permute-decoder}
\hspace*{0.02in} {\bf Input:}
The received signal $\mathbf{y} = \{y_i, i=1\cdots N\}$;\\
\hspace*{0.02in} {\bf Output:}
The recovered codeword $\hat{\mathbf{x}} = \{\hat{x_i}, i = 1 \cdots N\}$;\\
\begin{algorithmic}[1]
\STATE Initilize $L_{chan, i} \triangleq \frac{2y_i}{\sigma^2}$ for $i=1\cdots N$; $T_{\pi,i,j}^{0}=0 \ \forall i,j$; $\Lambda = \cal G$; \\
\FOR{Iterations: $t=1 \cdots t_{\max}$}
\STATE Select decoding graph: $\Lambda = ~ (\Lambda ==\cal G)~?~{\cal G}_\pi~:~{\cal G}$;
\IF {$\Lambda$ is ${\cal G}$}
	\FOR{Inner component codes: $i=1 \cdots \sqrt{N}$ (in parallel)}
	\STATE  $L_{i,j}^t = L_{chan, i + (j-1)\sqrt{N}} + {\alpha}_t T_{\pi,i,j}^{t-1}$ for $j=1 \cdots \sqrt{N}$;  
	\STATE $T_{i,j=1 \cdots \sqrt{N}}^{t}=SoftDecoder(L_{i,j=1 \cdots \sqrt{N}}^t)$;
	\ENDFOR
\ELSE
	\FOR{Inner component codes: $i=1 \cdots \sqrt{N}$ (in parallel)}
	\STATE   $L_{\pi,j,i}^t = L_{chan, (i-1)\sqrt{N}+j} + {\alpha}_t T_{j,i}^{t-1}$ for $j=1 \cdots \sqrt{N}$;  
	\STATE $T_{\pi,j=1 \cdots \sqrt{N}, i}^{t}=SoftDecoder(L_{\pi,j=1 \cdots \sqrt{N},i}^t)$;
	\ENDFOR
\ENDIF
\ENDFOR
\FOR{Inner component codes: $i=1 \cdots \sqrt{N}$}
\STATE  $\hat{x}_{i + (j-1)\sqrt{N}} = (L_{chan, i + (j-1)\sqrt{N}}+T_{i,j}^{t_{max}}+T_{\pi,i,j}^{t_{max}}< 0)$, for $j=1 \cdots \sqrt{N} $;
\ENDFOR

\end{algorithmic}
\end{algorithm}

Denote by $\cal G$ the original decoding graph consisting of $n$ stages. There are $n!$ equivalent stage-permuted graphs \cite{scpermutation1}.
Among them, we choose the permuted graph ${\cal G}_\pi$ with stage permutation $\pi\{1,2,...n\} = \{n/2+1, ...n, 1,...,n/2\}$.
This results into a swap between the inner and outer code parts of the original decoding graph $\cal G$ (see Fig.~\ref{permutation}).
Because the inner code part of ${\cal G}_\pi$ is the outer code part of $\cal G$, by decoding the inner code parts of $\cal G$ and ${\cal G}_\pi$, full information (parity check functions) about $\cal G$ is exploited.

The decoding algorithm iterates by decoding the two graphs $\cal G$ and ${\cal G}_\pi$ \emph{alternately} (line 3 in Algorithm~\ref{alg:stage-permute-decoder}) as follows.
For decoding graph $\cal G$ (resp. ${\cal G}_\pi$), the $j$-th code bit of the $i$-th inner component code is denoted by $x(i,j)$ (resp. $x_\pi(j,i)$).
Take the non-permuted graph $\cal G$ for example, $L_{i,j}^t$ is the log likelihood ratio (LLR) of the code bit $x(i,j)$ in the $t$-th iteration.
Specifically, $L_{i,j}^t$ is the sum of channel LLR $L_{chan,i + (j-1)\sqrt{N}}$ and the soft extrinsic information $T_{\pi,i,j}^{t-1}$ from the previous decoding iteration (line 6).
Following the method in \cite{product}, the extrinsic information is multiplied by a damping factor ${\alpha}_t$ to improve performance.
The $i$-th soft-output component decoder, denoted by $SoftDecoder()$, takes $L_{i,j}^t$, $j = 1,2,\cdots,\sqrt{N}$ as input, and generates extrinsic information $T_{i,j}^{t}$, $j = 1 \cdots \sqrt{N}$ as output (line 7).
There are $\sqrt{N}$ inner component codes in each decoding graph and the $\sqrt{N}$ component decoders can be implemented in parallel.
After $t_{\max}$ iterations, the algorithm outputs the hard decisions of combined LLRs as recovered code bits.

The decoding algorithm can exploit the parity check functions from both graphs.
During decoding each graph, a $\sqrt N$-times parallelism gain is obtained thanks to the fully parallel decoding of inner component codes.
Extrinsic information output of these component codes is iteratively exchanged until reaching a consensus.
We will show next that various types of soft-output decoders can be implemented to trade off between performance and decoding latency.

\subsection{Soft-output decoders for inner codes}
Since each inner component code is itself a short $G_N$-coset code, it is feasible to adopt low complexity SC-based decoders\cite{SCL,CASCL} as follows.
\begin{itemize}
  \item \emph{Type-1:} {\bf Soft-output SC list decoder} provides the best performance but has the highest complexity and latency. A Chase-like algorithm \cite{product} is used to generate soft LLR estimation from the decoding paths.
  \item \emph{Type-2:} {\bf Soft-output SC permutation list decoder} runs several permuted SC decoders in parallel. These independent SC decoders requires no sorting, thus is faster than the SC list decoder. The same Chase-like soft LLR generation method is used.
  \item \emph{Type-3:} {\bf Soft cancellation decoder} \cite{SCAN} can directly output soft LLRs. It has the smallest complexity and latency.
\end{itemize}

The above SC-based component decoders imply that the inner component codes could be constructed as polar codes.
Specifically, we may decode the inner codes using SC list $L$ decoders (Type-1).
A recently proposed SC permutation list decoding method \cite{scpermutationreed,scpermutation} can also be adopted (Type-2).
Specifically, for each inner component code, we perform SC decoding \emph{in parallel} on $L$ permuted codewords.
It does not involve any sorting operations among the list paths, therefore can further improve the parallelism and reduce latency within each component code.
Either way, for the $i$-th inner code, we obtain $L$ estimated codewords denoted by $\textbf{x}_i^l= \{{\hat x}^{l}_{i,1},{\hat x}^{l}_{i,2},...,{\hat x}^{l}_{i,\sqrt{N}}\}$ for $l\in\{1,2,...,L\}$.
The decoding results are then used to calculate the extrinsic information about code bits as follows.
For each estimated codeword, a mean square error is calculated as follows:
\begin{equation}\small\label{metric_for_sequence}
\begin{aligned}
M_i^{l} = \sum_{j=1}^{\sqrt{N}} (\frac{\sigma^2L_{i,j}^t}{2}  - (1-2{\hat x}_{i,j}^l))^2
\end{aligned}.
\end{equation}
Then, inspired by Chase decoding\cite{product}, we take $M_i^{l}$ as the path metric to calculate the soft output $T_{i,j}^t$:
\begin{equation}\small\label{llr}
\begin{aligned}
T_{i,j}^t = \frac{\min_{\{l|{\hat x}^{l}_{i,j}=1\}}M^{l}_i  - \min_{\{l|{\hat x}^{l}_{i,j}=0\}}M^{l}_i}{2\sigma^2}
\end{aligned}.
\end{equation}
When the decoded bits ${\hat x}^{l}_{i,j}$ are the same in all the $L$ estimated codewords, it means that the bit value is very likely to be correct and thus the soft output output $T_{i,j}^t$ is simply set to a large value.

The soft cancellation decoder (Type-3), proposed in \cite{SCAN}, can also be adopted as the inner code decoder.
This algorithm can produce (extrinsic) reliability information for the estimated code bits in a much simpler way.
Specifically, only soft decisions are made and propagated in the factor graph following the same scheduling as an SC decoder.
It does not require to maintain $L$ list paths as the SC list and SC permutation list decoders do.
Therefore, the soft cancellation decoder has a latency similar to an SC decoder, and requires much smaller memory than the previous two list decoders.

\section{code construction principle for the stage-permuted decoder}\label{section_encoding}
\label{fig_informationset}
\begin{figure}[]
\centering
\includegraphics[width=3.5in]{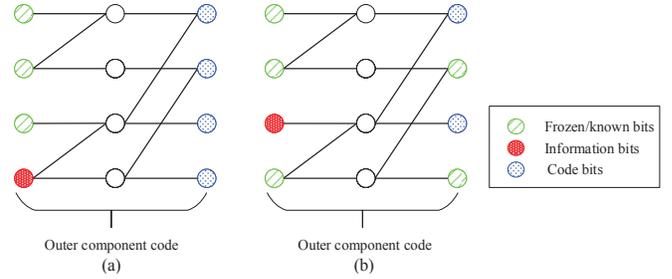}
\caption{The optimal code construction under stage-permuted decoding is different from both polar and RM constructions.
For a length-4 outer component code, according to polar or RM principle, the last bit should be the information bit, as shown in (a).
As a result, all code bits are unknown and then regarded as information bits while decoding the inner codes. Alternatively, if the third bit is the information bit, as shown in (b), two code bits become known bits. This reduces the code rate of inner codes.}
\label{rmvsnewconstruction}
\end{figure}
As discussed, the extrinsic information from the inner code decoders are exchanged between the two graphs during the decoding.
Therefore, the decoding performance of inner codes is essential for the overall performance.
This requires specific code constructions (different from both polar and RM principles), as elaborated in the following example.

Consider a length-16 $G_N$-coset code consisting of four length-4 component codes. Assume that an outer component code has one information bit.
As shown in Fig.~\ref{rmvsnewconstruction}(a), either RM or polar principle would request the last bit to be selected as the information bit\cite{po}.
As a result, all code bits are unknown and thus regarded as information bits of the inner component codes.
In other words, the inner code rates become higher, leading to poorer performance of the corresponding inner component decoders.

In contrast, consider the case that the third bit be the information bit. As shown in Fig.~\ref{rmvsnewconstruction}(b), two of the code bits are known bits (set to $0$).
For the inner component codes, these two code bits become frozen bits and thus reduce the code rate of inner codes.

This example demonstrates the disadvantage of RM/polar principle, and illustrates the heuristic of our code construction algorithm. In the following, we propose an information set selection rule that maximally reduces the inner code rates.

\subsection{Choose information set for $K=k^2$}
The construction involves two steps:

\begin{enumerate}
\item \emph{$(\sqrt{N}, k)$ component codes:} since we proposed SC-based decoders for the inner codes, the  ideal construction is a $(\sqrt{N},k)$ short polar code.
Denote by $P=[p_1, p_2,...,p_{\sqrt{N}}]$ the information vector:
\begin{equation}
p_i=\left\{
\begin{array}{rcl}
1 & & {\text{The } i\text{-th bit is an information bit.}}\\
0 & & {\text{The } i\text{-th bit is a frozen bit.}}\\
\end{array} \right.
\end{equation}

\item \emph{$G_N$-coset codes}: denote by $I$ the information vector of the stage-permuted $G_N$-coset codes. It can be derived from $P$ as follows:
\begin{equation}
\small\label{IL}
\begin{aligned}
I = P \otimes P
\end{aligned}.
\end{equation}

\end{enumerate}

For example, consider a $(16, 9)$ stage-permuted $G_N$-coset code construction.
In the first step, we construct a $(4, 3)$ polar code. The information vector $P$ is as follows:
\begin{equation}\small\label{s1}
\begin{aligned}
    P=[0~1~1~1]
\end{aligned}.
\end{equation}
Then, the information vector $I$ of the stage-permuted $G_N$-coset code is obtained as follows:
\begin{equation}\small\label{s2}
\begin{aligned}
I = P \otimes P = [0~0~0~0~0~1~1~1~0~1~1~1~0~1~1~1]\\
\end{aligned}.
\end{equation}

Compared with polar and RM constructions, this code construction reduces the perceived coding rates at the inner components decoders.
All the inner component codes have the same information vector $P$. Since $P$ is constructed by the polar principle, inner component codes are efficiently decoded by SC-based decoders.
\label{fig_informationset}
\begin{figure}[]
\centering
\includegraphics[width=3.5in]{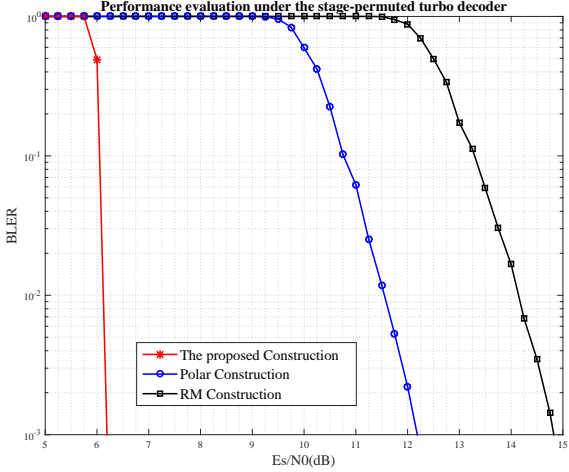}
\caption{Under stage-permuted decoding, the proposed stage-permuted $G_N$-coset code achieves significantly better performance than polar and RM constructions. $N=65536$ and $K=57121$.}
\label{fig_s3}
\end{figure}

With either polar or RM principle to construct a $G_N$-coset code, several information bits would be allocated to consecutive bit positions at the ending part of an information block. This would significantly degrade the performance if a stage-permuted turbo-like decoder is applied, because they might as well yield all-rate-1 inner component codes. Fig.~\ref{fig_s3} provides a numerical simulation result to support this assertion. As expected, our code construction principle to avoid higher coding rate for inner codes generates better performance than both polar and RM ones if the stage-permuted turbo-like decoder is applied.

\subsection{General code construction}
To construct an $(N, K)$ code, we first construct an $(N, K_1)$ stage-permuted $G_N$-coset code according to the previous subsection, where $K_1\triangleq\lceil{\sqrt{K}}\rceil^2$ is the first square number larger than $K$.
Then, among the $K_1$ information bit positions, we additionally freeze $K_1-K$ bit positions.

Unlike the polar principle that would freeze the $K_1-K$ least reliable bit positions, our heuristic construction reduces the code rates for the inner codes in an iterative way.
In each iteration, we freeze one bit position that would reduce the inner code rate.
This incremental freezing is performed alternately on the original decoding graph $\cal G$ and the stage-permuted decoding graph ${\cal G}_\pi$ until $K$ information bit positions are left.
The details are given in Algorithm~\ref{alg:Framwork}, Algorithm~\ref{alg:FindMHIBs} and Algorithm~\ref{alg:rhs}, and explained as follows.

\begin{algorithm}[htb]
\caption{A successive freezing algorithm.}
\label{alg:Framwork}
\hspace*{0.02in} {\bf Input:}
Information vector $I$;\\
\hspace*{0.02in} {\bf Output:}
Newly-constructed information vector $I_o$;\\
\begin{algorithmic}[1]
\STATE $N = \text{length}(I)$, $K_1 = \sum{I}$, $I_o = I$
\FOR{$i=1$; $i\leq K_1 - K$; $i=i+1$}
\IF {$i$ is odd}
\STATE $j$ = SelectOneBitPositionToFreeze($I_o$, ${\cal G}$);
\ELSE
\STATE $j$ = SelectOneBitPositionToFreeze($I_o$, ${\cal G}_\pi$);
\ENDIF
\STATE $I_o(j) = 0$;
\ENDFOR
\end{algorithmic}
\end{algorithm}

\begin{figure}[]
\centering
\includegraphics[width=3.5in]{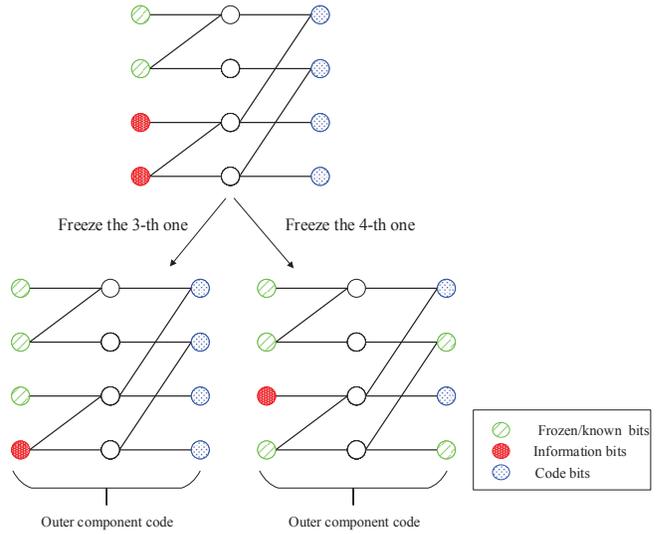}
\caption{ For a length-4 outer component code with last two bit positions as information set, the last bit position is an RRBP while the third one is not.
After freezing the last bit position, two code bits become frozen bits (as shown in the right graph), which reduce the code rate. In contrast, if the third one is freezed, all the code bits are unknown (as shown in the left graph).}
\label{fig_helpful_bits}
\end{figure}

Firstly, we narrow down to the rate-reducing bit positions (RRBPs), which have the following property (also illustrated in Fig.~\ref{fig_helpful_bits}).
When an RRBP $u_i$ is freezed, at least one of the corresponding outer component code bits becomes a known bit, denoted by $x_i$.
From the inner code's perspective, bit $x_i$ is decoded as a frozen bit and thus reduces the code rate.

Secondly, we only freeze one bit position among the RRBPs in each iteration.
When there are multiple RRBPs, we choose one RRBP $u_i$, such that the resultant frozen bit $x_i$ in the inner component codes has the least reliability $r_i$.
As a result, the remaining information set of each inner component code is still optimal according to the polar principle, and thus can be efficiently decoded by SC-based decoders.
The details are given in Algorithm~\ref{alg:FindMHIBs} and  Algorithm~\ref{alg:rhs}.

With the general algorithm, $G_N$-coset codes with arbitrary code rates can be constructed.
The proposed method is designed such that the performance under the stage-permuted turbo-like decoder is optimized.
The heuristic is to reduce the coding rate of the inner codes, as well as maximally preserving their sub-channel reliabilities.

\begin{algorithm}[htb]
\caption{SelectOneBitPositionToFreeze($I$, $\Lambda$)}
\label{alg:FindMHIBs}
\hspace*{0.02in} {\bf Input:}
Information vector $I$, decoding graph $\Lambda$;\\
\hspace*{0.02in} {\bf Output:}
The index of the bit to freeze;\\
\begin{algorithmic}[1]
\STATE $N = \text{length}(I)$, $\Phi = []$;
\STATE $\Gamma$ = InnerInformationVector(I, $\Lambda$);
\FOR{$i=1$; $i\leq N$; $i=i+1$ }
\IF {$I(i)$ is $1$}
\STATE $I_i = I$, $I_i(i)=0$;
\STATE $\Gamma_i$ = InnerInformationVector($I_i$, $\Lambda$);
    \IF {$\Gamma_i$ is not equal to $\Gamma$}
    \STATE $\eta_i = \min\{index( \Gamma_i ~ !=~\Gamma )\}$;
    \IF {$\Lambda$ is ${\cal G}$}
    \STATE   $\eta_i = \text{int} (\frac{\eta_i-1}{\sqrt{N}})+1$;
    \ELSE
    \STATE   $\eta_i = (\eta_i-1)\%\sqrt{N}+1$;
    \ENDIF
    \STATE $\Phi$.append($\{i, r_{\eta_i}\}$);
    \ENDIF
\ENDIF
\ENDFOR
\RETURN $\argmin_i\{r_{\eta_i} \in \Phi \}$.
\end{algorithmic}
\end{algorithm}

\begin{algorithm}[htb]
\caption{InnerInformationVector($I$, $\Lambda$)}
\label{alg:rhs}
\hspace*{0.02in} {\bf Input:}
Information vector $I$, decoding graph $\Lambda$;
\\\hspace*{0.02in} {\bf Output:}
Information vector $I_o$;\\
\begin{algorithmic}[1]
\STATE $I_o = I$
\IF{$\Lambda$ is ${\cal G}$}
\STATE $i_s \Leftarrow 1$;
\ELSE
\STATE $i_s \Leftarrow \frac{\log_2(N)}{2} + 1$;
\ENDIF
\FOR{$i=i_s$; $i<\frac{1}{2}\log_2(N)+i_s$; $i=i+1$ }
\STATE ${\cal N} = 2^{i}$, $\triangle = \frac{\cal N}{2}$, ${\cal M} = \frac{N}{\cal N}$;
\FOR{$m=0$; $m<{\cal M}$; $m=m+1$ }
\FOR{$z=1$; $z\leq \triangle$; $z=z+1$}
\IF {$I_o(z+\triangle+m*{\cal N})$ is $1$}
\STATE $I_o(z+m*{\cal N})=1$;
\ENDIF
\ENDFOR
\ENDFOR
\ENDFOR
\RETURN $I_o$.
\end{algorithmic}
\end{algorithm}

\section{Performance evaluation}\label{section_simulation}
In this section, we evaluate the performances and parallelism of several coding schemes. The coded symbols are modulated with binary phase-shift keying (BPSK) modulation and then transmitted over an additive white Gaussian noise (AWGN) channel.

\label{fig_informationset}
\begin{figure}[]
\centering
\includegraphics[width=3.5in]{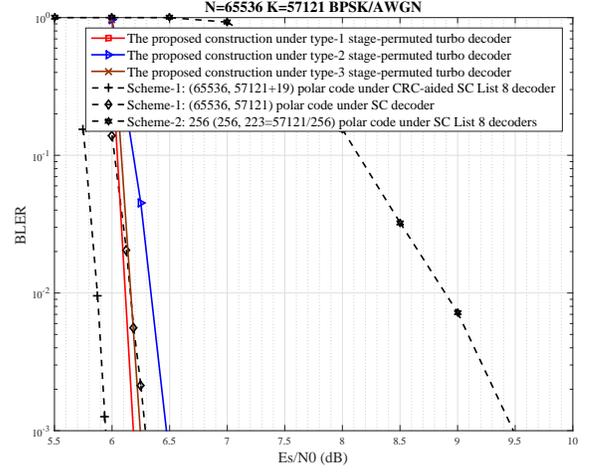}
\caption{BLER performance in the case with a square number of information bits. Compared to Scheme-2, our scheme achieves significantly better performance.
Compared to Scheme-1, our scheme achieves comparable error correction performance. }
\label{fig_s1}
\end{figure}

\label{fig_informationset}
\begin{figure}[]
\centering
\includegraphics[width=3.5in]{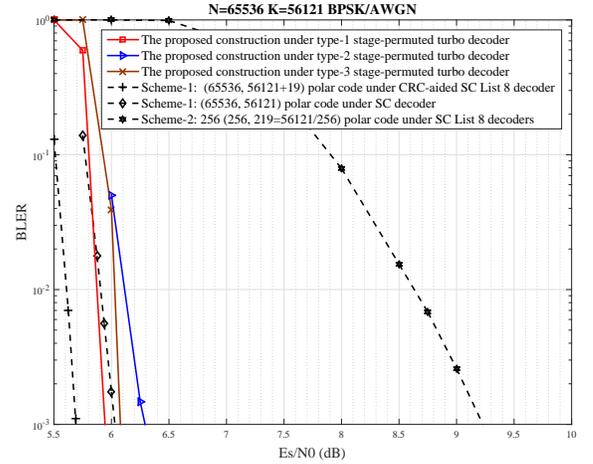}
\caption{BLER performance in the case with a general number of information bits. Compared to Scheme-2, our scheme achieves significantly better performance.
Compared to Scheme-1, our scheme achieves comparable error correction performance. }
\label{fig_s2}
\end{figure}

The proposed $G_N$-coset codes are decoded by the stage-permuted decoding algorithms with 8 iterations.
During the decoding of inner codes, the SC list 8 decoding algorithm (Type-1), the SC permutation algorithm with 8 permutations (Type-2), the soft successive cancellation algorithm (Type-3) are evaluated.
In the simulations, the damping factors are set as follows.  For the the SC list 8 decoding algorithm (Type-1) and SC permutation algorithm (Type-2), the damping factors $\alpha = [0.3,0.3,0.4,0.5,0.6,0.7,0.8,1]$. For the soft successive cancellation algorithm (Type-3),
the damping factors $\alpha = [3/8, 3/8, 3/8, 3/8,3/8, 3/8,3/8, 4/8]$.

We first evaluate the code construction with a square number of information bits. In the simulation, we construct $(65536, 57121=239^2)$ stage-permuted $G_N$-coset codes and then
decode them with all types of decoding algorithms. 
Then, we evaluate the general code construction with $N = 65536$ and $K = 56121$.

Polar codes with different configurations are compared as benchmarks.
In Scheme-1, the same number of information bits are encoded to a length-$65536$ polar code. This long code configuration obtains more coding gain but incurs larger decoding latency.
In Scheme-2, 256 length-$256$ short polar codes are encoded and decoded in parallel. This short code configuration exhibits a similar degree of parallelism to ours, but suffers from performance loss.
The polar codes are decoded by SC decoders and CRC-aided (CA with 19 CRC bits) SC list 8 decoders.

The block error rate (BLER) performances are provided in Fig.~\ref{fig_s1} and Fig.~\ref{fig_s2}.
Compared to Scheme-2, our scheme achieves significantly better performance.
Compared to Scheme-1, our scheme achieves comparable error correction performance.
However, the decoding latency of our scheme is much smaller than Scheme-1, as discussed below.

The decoding latency is evaluated with an ASIC implementation in a 16nm TSMC FinFET technology~\cite{Jiajie}.
The required time steps of these coding schemes are given in Table~\ref{tab:latency}.
It demonstrates that our scheme can significantly reduce the decoding latency thanks to the high degree of parallelism.
Therefore, the proposed $G_N$-coset codes possess the benefits of both coding gain (comparable to that of Scheme-1) and parallelism (comparable to that of Scheme-2).

Finally, we compare the proposed three types of soft-output component decoders. With the Type-1 (SC list) component decoder, it achieves better decoding performance with more time steps. On the contrary, with the Type-2 (SC permutation list) and Type-3 (soft cancellation) component decoders, the required time steps can be reduced significantly.
This only comes at a cost of 0.3 and 0.1 dB performance loss, respectively. The diverse choices of component decoders provide a flexible trade-off between performance and decoding latency to meet the requirements of various communication scenarios.

\begin{table}[!htbp]
\caption{A comparison of the required time steps between the proposed coding schemes and polar codes.}
\label{tab:latency}
\centering
   \begin{threeparttable}
\begin{tabular}{|c|c|c|c|c|} 
\hline
Scheme&N&Rate&Time steps&\small{Parallelism}\\
\hline
\multirow{8}*{\tabincell{c}{Type-1:\\Soft-output SC list\\as inner decoder}}&	\multirow{8}*{65536}	&\multirow{4}*{\tabincell{c}{$0.8716$\\ $(\frac{57121}{65536})$ }}&20160	&16\\
\cline{4-5}
	&	&	&10080&	32\\
\cline{4-5}
	&	&	&5040&	64\\
\cline{4-5}
	&	&	&2520&	128\\
\cline{3-5}
	&	&	\multirow{4}*{\tabincell{c}{$0.8563$\\ $(\frac{56121}{65536})$ }}&	20032&	16\\
\cline{4-5}
&	&	&10016&	32\\
\cline{4-5}
&	&	&5008&	64\\
\cline{4-5}
&	&	&2504&	128\\
\hline
\multirow{8}*{\tabincell{c}{Type-2:\\Soft-output SC\\permutation list\\as inner decoder}}&	\multirow{8}*{65536}	&\multirow{4}*{0.8716}&	4736	&16\\
\cline{4-5}
&	&	&2368	&32\\
\cline{4-5}
&	&	&1184&	64\\
\cline{4-5}
&	&	&592	&128\\
\cline{3-5}
&	&		\multirow{4}*{0.8563}&	4864&	16\\
\cline{4-5}
&	&	&2432&	32\\
\cline{4-5}
&	&	&1216&	64\\
\cline{4-5}
&	&	&608&128\\
\hline
\multirow{8}*{\tabincell{c}{Type-3:\\Soft cancellation\\as inner decoder}}&	\multirow{8}*{65536}	&\multirow{4}*{0.8716}&	10272 &16\\
\cline{4-5}
&	&	&5136	&32\\
\cline{4-5}
&	&	&2568&	64\\
\cline{4-5}
&	&	&1284	&128\\
\cline{4-5}
&	&	&642  &256\\
\cline{3-5}
&	&		\multirow{4}*{0.8563}&	11008&	16\\
\cline{4-5}
&	&	&5504&	32\\
\cline{4-5}
&	&	&2752&	64\\
\cline{4-5}
&	&	&1376 &128\\
\cline{4-5}
&	&	&688 &256\\
\hline
\multirow{2}*{\tabincell{c}{Polar\footnote{This is evaluated with the double-package mode closed\cite{Jiajie}.} \\ CA SC List 8}}&\multirow{2}*{65536}&	$0.8716$	&93097&	1\\
\cline{3-5}
& &$0.8563$&	93477	&1\\
\hline
\multirow{2}*{\tabincell{c}{Polar\\ SC}}&	\multirow{2}*{65536}	&0.8716 &	13146	&1\\
\cline{3-5}
&	&0.8563	&13282&	1\\
\hline
\end{tabular}
      \begin{tablenotes}
        \footnotesize
        \item[1] This is evaluated in our ASIC implementation \cite{Jiajie} with the double-package mode turned off.
      \end{tablenotes}
    \end{threeparttable}
\end{table}

\section{Conclusion}\label{section_conclusion}
We study the construction of $G_N$-coset codes decoded by a stage-permuted turbo-like decoding algorithm.
Through stage permutation, the decoding algorithm can exploit the parity check functions from multiple equivalent factor graphs.
Since only the inner code parts are decoded (in parallel) and the outer code processing is avoided, the decoding algorithm exhibits a higher degree of parallelism.
Based on this decoding algorithm, we propose a new $G_N$-coset code construction for arbitrary information lengths and coding rates. 
The novel encoder-decoder framework is evaluated in terms of both performance and decoding latency.
The simulations suggest that the constructed $G_N$-coset codes achieve comparable error correction performance to polar codes of the same length.
The ASIC implementation evaluation verifies that the stage-permuted turbo-like decoding algorithm has a much lower decoding latency.

\bibliographystyle{IEEEtran}

\end{document}